# Large-sensitive-area superconducting nanowire single-photon detector at 850 nm with high detection efficiency

Hao Li, Lu Zhang, Lixing You,* Xiaoyan Yang, Weijun Zhang, Xiaoyu Liu, Sijing Chen, Zhen Wang, and Xiaoming Xie

*State Key Laboratory of Functional Materials for Informatics, Shanghai Institute of Microsystem and Information Technology (SIMIT), Chinese Academy of Sciences, 865 Changning Rd., Shanghai 200050, China*

*lxyou@mail.sim.ac.cn

**Abstract**: Satellite–ground quantum communication requires single-photon detectors of 850-nm wavelength with both high detection efficiency and large sensitive area. We developed superconducting nanowire single-photon detectors (SNSPDs) on one-dimensional photonic crystals, which acted as optical cavities to enhance the optical absorption, with a sensitive-area diameter of 50 μm. The fabricated multimode fiber coupled NbN SNSPDs exhibited a maximum system detection efficiency (DE) of up to 82% and a DE of 78% at a dark count rate of 100 Hz at 850-nm wavelength as well as a system jitter of 105 ps.

## 1. Introduction

Superconducting nanowire single-photon detectors (SNSPDs) have received considerable attention because of their high detection efficiency (DE), low timing jitter, high count rate, and low dark count rate (DCR). In the past decade, the optical absorptions have been significantly enhanced due to integration of optical cavities that successfully increased the DE to over 70% and even to 90% at a wavelength of 1550 nm[1-7], enabling numerous impressive applications such as long-distance quantum key distribution (QKD) [8, 9], space-ground laser communication[10], depth imaging[11, 12], and on-chip characterization of nanophotonic circuits[5]. However, many applications still require single-photon detectors (SPDs) with high DE at wavelengths other than 1550 nm. In addition, a large sensitive area is another prerequisite for some applications because free space coupling is typically used, such as for satellite–ground QKD at 850 nm[13, 14]; satellite laser ranging designed at 1064, 850, or 532 nm[15, 16]; fluorescence spectroscopy at 635 nm[17]; and many other free-space coupling applications. Recently, some key experimental verifications have been conducted toward satellite–ground QKD, and the Chinese satellite is planned to launch in 2016[14, 18]. The SPD used in the satellite–ground QKD project is a PerkinElmer SPCM-AQRH-16 with multimode-fiber–coupled (MMF-coupled) avalanche photodiodes (APDs) [14], which has a DE of over 45% with a sensitive area of D = 180 μm and DCR of 25 Hz at 850 nm.

Previous work on large-sensitive-area SNSPD (D = 35 μm) was reported with the maximal DE over 70% at visible wavelengths (~50% at 850-nm wavelength)[19]. In this work, we first demonstrated SNSPDs with high performance at the wavelength of 850 nm. To enhance the absorption at 850 nm, a specific substrate with a one-dimensional (1-D) photonic crystal (PC) comprising dielectric films are used. The sensitive area diameter is 50 μm, which to the best of our knowledge, has the largest sensitive area for a single pixel SNSPD[20-23]. By optimizing the linewidth and the spacing, the best device showed a maximal DE of 82% and DE of 78% at DCR of 100 Hz, which have prospective applications in free space satellite-ground quantum communication.

**2. Device design and fabrication**

Our devices were designed on the basis of a 1-D PC to enhance the absorption of the nanowire, as illustrated in Fig. 1(a). In the 1-D PC, the periodic structure will result in a photonic band gap, in which the photons are forbidden from propagating through and are reflected back. Thus, the optical absorption of the nanowire fabricated on the PC can be enhanced. Common PC materials include $Ta_2O_5$ and $SiO_2$ bilayers with refractive indices of $n_{Ta_2O_5}$ = 2.05 and $n_{SiO_2}$ = 1.47, respectively, which were measured using spectroscopic ellipsometer at a wavelength of $\lambda_0$ = 850 nm at room temperature. For a resonance absorptance around the target wavelength, the thicknesses of the layers are designed to be a quarter wavelength of $\lambda_0$, $L_1$ = 104 nm for $Ta_2O_5$ and $L_2$ = 145 nm for $SiO_2$. Theoretically, we can achieve a reflectivity infinitely close to unity by increasing the number of the bilayers. Here, we selected 13 bilayers for high reflectivity and acceptable fabrication complexity. Similar PC structures have also been applied to improve the absorption of the SNSPD for other wavelengths[19, 24].

Experimentally, the 13 periodic $SiO_2/Ta_2O_5$ bilayers are alternatively deposited one layer at a time on a Si substrate using ion beam sputtering, with the film thickness monitored to ensure adherence to the designed layer thickness. The measured surface roughness of the PC is less than 3 Å, and the measured reflectivity at 850 nm is more than 99%. Then, an ultrathin NbN film is deposited on the PC substrate at room temperature using reactive DC magnetron sputtering in a mixture of Ar and $N_2$ gases (partial pressures of 79% and 21%, respectively). The thickness of the film was controlled by the sputtering time and verified using X-ray reflectometry. NbN was selected because NbN SNSPD exhibits excellent performance at 2.1 K using a commercial Gifford–McMahon cryocooler. The film was patterned into a meandered nanowire structure by electron beam lithography (EBL) using a positive-tone polymethyl methacrylate electron-beam resist and was reactively etched in $CF_4$ plasma. Proximity-effect correction was not carried out for EBL process and the nominal linewidth of the nanowire were estimated according to SEM images and the layout parameters. Then, a 50-Ω-matched coplanar waveguide was formed using ultraviolet lithography and reactive ion etching. A cross-sectional transmission electron microscopy (TEM) image of the nanowire as well as the PC substrate is presented in Fig. 1(b). Figure 1(c) presents a scanning electron microscopy (SEM) image of the active area with a diameter of 50 μm. The magnified SEM image for the nanowire is shown in Fig. 1(d), indicating the width/pitch of 120 nm/200 nm.

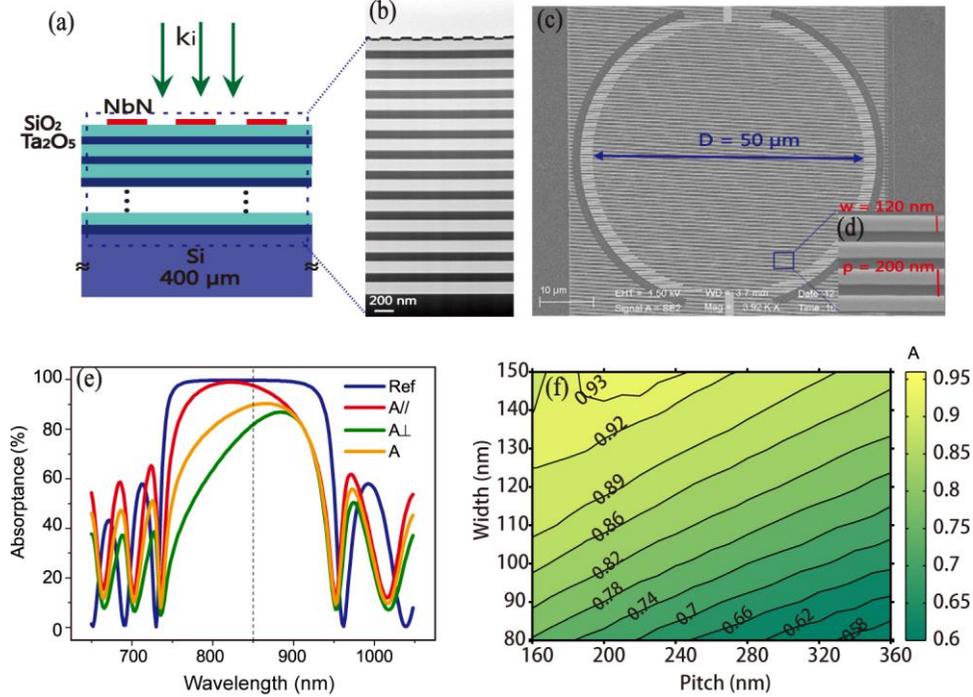

Fig. 1 (a) Schematic of the SNSPD based on PC substrate. (b) TEM image of a cross section of the nanowire on the PC substrate. The PC structure was formed by multiple layers of alternating $Ta_2O_5$ and $SiO_2$ layers on a Si substrate. (c) SEM of the active area with a diameter of 50 μm; (d) Magnified SEM image of nanowire with width w = 120 nm and pitch p = 200 nm. (e) Reflectivity of the PC based on Si substrate (blue) consisting of 13 bilayers composed of $Ta_2O_5$ and $SiO_2$, absorptance of the nanowire for parallel polarization waves $A_{//}$ (red), perpendicular polarization waves $A_{\perp}$ (green) and the average absorptance $A = (A_{//} + A_{\perp})/2$ (yellow) for normal incidence calculated using the RCWA method; (f) The calculated average absorptance A versus pitch and width of nanowire with thickness of 6.5 nm.

To quantitatively investigate the optical absorption of SNSPDs, an electromagnetic (EM) simulation is required [1, 4, 25]. We performed an EM simulation using the rigorous coupled-wave analysis (RCWA) method, in which the EM field in the periodic grating region (nanowire region) is expanded into a sum of spatial harmonics on the basis of Floquet's theorem. The refractive index of NbN, $n_{NbN} = 4.53 + 4.57i$, at approximately 850 nm was obtained using a spectroscopic ellipsometer, and the refractive index and thickness of the Si substrate were $n_{Si} = 3.46$ and $l = 400$ μm, respectively. The polarization dependence of the SNSPD may limit the DE when an SNSPD is coupled with an MMF because the polarization in MMFs is difficult to control because of the coupling of different propagating modes caused by the fiber imperfections, such as index inhomogeneity, core ellipticity and eccentricity, and bends. Therefore, in the simulation, we define the average absorptance $A = (A_{//} + A_{\perp})/2$ to represent the average absorptance of our structure, where $A_{//}$ and $A_{\perp}$ are the calculated absorptance for the parallel and perpendicular polarization plane waves, respectively. Figure 1(e) shows the calculated absorptance $A_{//}$, $A_{\perp}$, and A for normal incidence. One can see the absorption peaks around the target wavelength 850 nm for both polarizations resulting from the resonance effect of structure. An average absorptance A of 89.4% can be achieved with this structure at 850-nm wavelength. The thickness, pitch, and width of the NbN nanowire were selected to be 6.5, 200, and 120 nm in the calculation, respectively. The detailed nanowire pitch and width dependence of the average absorptance A were also calculated, and the results are presented in Fig. 1(f), which helped us to optimize the geometric parameters of the nanowires. We noticed that a higher filling ratio was beneficial to achieve a high absorption, which is different from the optical design for 1550 nm[4].

In our devices, the photons are directly guided to the SNSPD through a front-side aligned lensed fiber. The graded index lenses were spliced to the tip of the MMF with a 50-μm-diameter core. The incident light spot was focused to a minimal diameter $2w$ of about 30 μm (the beam waist) where the device was located. The fractional power for a Gaussian beam of the spot size $2w$ focusing on a centered circular nanowire area of diameter $2r$ is given by $P_c = \int 2\pi r I(r) dr = 1 - e^{\frac{-2r^2}{w^2}}$, where $I(r) = \frac{2}{\pi w^2} e^{-\frac{-2r^2}{w^2}}$ is the normalized radial field intensity variation. Thus, to obtain a good coupling, the diameter $2r$ of the nanowire area was selected to be 50 μm with $P_c = 0.996$.

## 3. Results and discussion

In our experiment, SNSPDs with various geometric parameters (film thickness, nanowire width and pitch) were fabricated to experimentally determine the optimized detection efficiency. The SNSPDs were packaged into a copper block, such that a lensed multi-mode fiber could be directly aligned with the sensitive area from the front-side. The focus of the fiber was located on the center of the meandered nanowire, which ensured a maximal optical coupling from the fiber to device. The package was mounted to the cold head of a two-stage Gifford–McMahon cryocooler with a working temperature of $2.100 \pm 0.005$ K. To measure DE, a pulsed laser was selected as the photon source. The single mode fiber (SMF) laser module was directly connected to the MMF aligned to the SNSPD. Thus only a small fraction of the core of the MM fiber was illuminated. No specific actions were taken to populate a large number of modes in the MMF. But this should not affect the measured DE since the diameter of SNSPD is large enough to ensure nearly unity coupling as described in the last paragraph and most of our devices showed a saturated DE behavior. The laser was then heavily attenuated to achieve a photon flux of $10^5$ photons/s. The DE of the detector was defined as DE = (OPR − DCR)/PR, where OPR is the output pulse rate of the SNSPD, as measured using a photon counter; DCR is the dark count rate when the laser is blocked; and PR is the total photon rate input to the system. At each bias current, an automated shutter in a variable attenuator blocked the laser light, and dark counts were collected for 10 s using a commercial counter. The light was then unblocked, and output photon counts were collected for another 10 s. Errors due to the calibration of the laser power were less than 3% given by the power meter, and the laser power fluctuation was less than 1%. Figure 2 shows the DE relations of the bias current for 15 different devices from three wafers with different NbN film thicknesses. Each wafer included 25 devices with different geometric parameters. The yield for different geometries varies on the fill factor. The yield for the devices with a small fill factor (≤120/250) is ~80%, while the yield for the devices with a large fill factor (≥120/200) is only ~20%. The best detectors were picked up for the same geometric parameter. All the curves in Fig. 2 are marked with the parameters of the nanowires (thickness/width/fill factor).

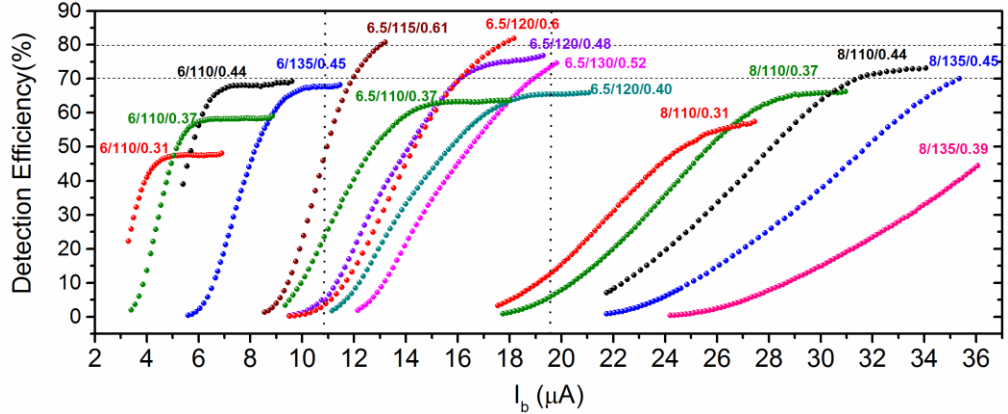

Fig. 2 DE as a function of the bias current for different devices. The curves are marked with the size of the nanowire as thickness/width/fill factor.

A few interesting results were obtained from these curves. Apparently, a high DE MMF coupled SNSPDs with a sensitive area D = 50 μm was obtained at 850-nm wavelength. The maximum system DE reaches 82% for the device with the geometric parameters 6.5/120/0.6 and 78% at DCR of 100 Hz, which is the highest value reported at the 850-nm wavelength. In comparison, the best commercial product Si APD (ID120-500-800 from ID Quantique Inc.) exhibits a DE of approximately 65% at 850 nm wavelength with a DCR down to 200 Hz. On the other hand, the measured DE of SNSPD is an averaged DE for the parallel and perpendicular polarization waves, which implies that the DE for the parallel polarization photons may be higher.

The curves in Fig. 2 have three groups corresponding to different film thickness. The switching currents of SNSPDs increased with the film thickness. For devices with the same thickness, the switching current tends to increase when the nanowire width was increased (see the 6/110 and 6/135 devices, 8/110 and 8/135 devices). It is reasonable and consistent with the previous report [26]. However, for the SNSPDs with same film thickness and nominal linewidth, we observed higher inflection currents (also higher switching currents) and higher DE for the higher filling factors (see the 6/110, 8/110 devices). The higher DE is consistent with the simulation result in Fig. 1(f). The higher inflection/switching currents can be explained by the EBL process without the proximity effect correction. When the filling factor of a meander is decreased (i.e. the pitch is increased), the nanowire linewidth tends to decrease with respect to the nominal width due to the proximity effect (over-exposure effect) in EBL process, which results in smaller switching currents and smaller inflection currents. Oppositely, many devices with a very high fill factor (for example, 6.5/120/0.67) had a relatively low switching current and DE (not shown here). This phenomenon may be explained by the current crowding effect at the 180 ° turn around at the end of straight nanowires[27, 28], even though the round corner design has already been applied in the structure. Even for the two detectors with maximal DE over 80% (6.5/115/0.61 and 6.5/120/0.6), we do not observe obvious saturation behavior, which indicates that the current crowding effect may also affect their switching current. Therefore, calculating an optimized high fill ratio without an evident current crowding effect is still necessary.

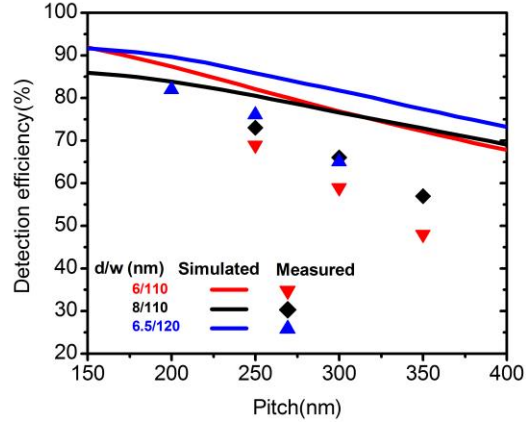

Fig. 3 Maximum measured DE extracted from Fig. 2 and simulated absorptance as functions of pitch. The simulated results are presented as solid lines, and the measured results are marked with diamonds and triangles. The sizes of the nanowire are marked as thickness/width (d/w).

The DE can be expressed as DE=$P_{couple} \times P_{abs} \times P_{pulse} \times (1-P_{loss})$, where $P_{couple}$ is the coupling efficiency between the incident light and the active area, $P_{abs}$ is the absorptance of nanowire, $P_{pulse}$ is the pulse generation probability of the nanowire and $P_{loss}$ represents any other optical loss in the system. Figure 3 shows the dependence of the calculated $P_{abs}$ and measured maximum DE on the pitch for three different film thicknesses and linewidths extracted from Fig. 2. One can see that the system DE decreases with the increase of the pitch, which is consistent with the calculation. The lower DE compared with the calculated absorption is mainly attributed to the system optical loss and the imperfect fabrication.

For the SNSPD with the highest DE of 82% (6.5/120/0.6), the DCR and jitter were characterized, and the results are presented in Fig. 4. The DE is approximately 78%/67% with DCR = 100 Hz/10 Hz obtained from Fig. 4(a). The timing jitter of the SNSPD was measured using the time-correlated single-photon counting (TCSPC) method[29]. At a bias current of 17.0 μA (DCR=100 Hz), the timing jitter defined by the full-width at half-maximum (FWHM) value of the histogram is 105.3 ps. The jitter value is higher than the jitter (~ 75.7 ps) of the SNSPD with a sensitive area of 35 μm [19]. One possible reason is the relative smaller switching current of our device (18.0 μA vs. 30.3 μA). One contribution of the system timing jitter is the jitter attributed to the signal noise ratio, which is proportional to $\sigma_n/k$, where $\sigma_n$ and $k$ are the root-mean-square noise of the signal and the slope of the rising edge of the signal [29]. $\sigma_n$ is mainly determined by the noise of the amplifier and $k$ roughly equals to the amplitude of the signal divided by the rising time. As a result, the higher bias current gives a higher signal amplitude, which corresponds to a lower timing jitter. On the other hand, the large sensitive area gives a larger kinetic inductance, which results in a longer rising time. The longer rising time causes a smaller $k$, thus producing a larger jitter.

Finally, to estimate the response speed, the measured oscilloscope persistence trace is presented in Fig. 4(c), which indicates a recovery time constant $\tau$ of approximately 160 ns. In addition, we measured an inductance $L_k$ of 8.0 μH by fitting the phase of the reflection coefficient using a network analyzer[23]. The inductance gives a recover time constant $\tau=L_k/R_L$=8.0 μH/50 Ω=160 ns, which agrees well with the oscilloscope persistence trace, where $R_L$ is the impedance of the load side.

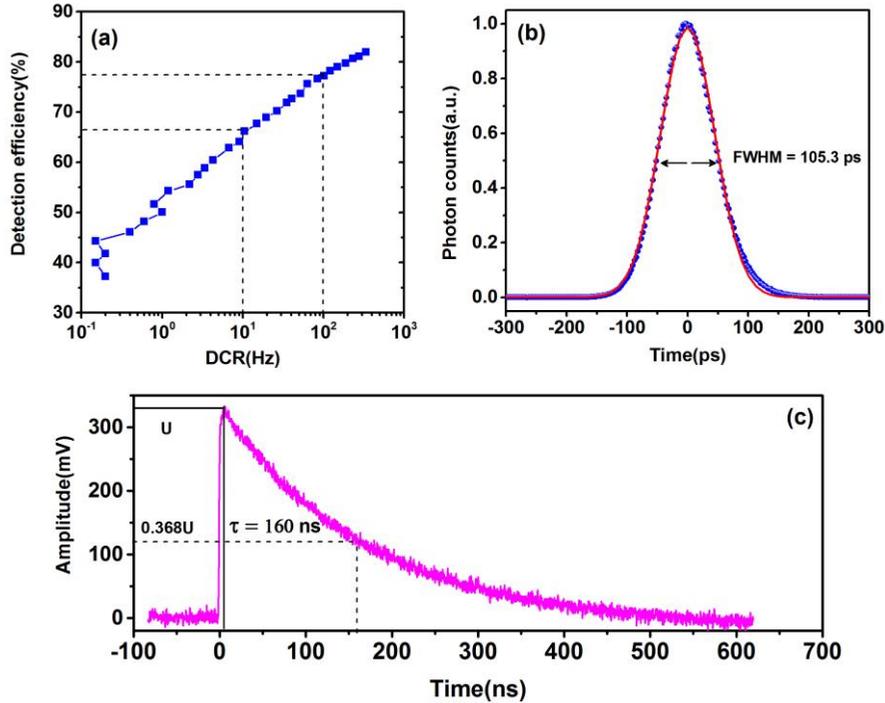

Fig. 4 (a) DE as a function of DCR for the SNSPD with the highest DE. The DE is approximately 78%/67% with DCR=100 Hz/10 Hz. (b) Histograms of the time-correlated photon counts measured at a wavelength of 1550 nm. The red lines are the fitted curves using the Gaussian distribution. (c) Oscilloscope persistence map of the response at a bias current of 17.0 μA.

### 4. Conclusions and outlook

We designed, fabricated, and characterized NbN SNSPDs at 850-nm wavelength on a PC substrate with a sensitive area diameter of 50 μm. The PC structure effectively acts as a cavity to enhance the absorption of incident photons. The MMF coupled SNSPDs exhibit the maximum system DE of up to 82% and system DE of 78% at DCR = 100 Hz, which is the highest DE reported for an SPD at 850 nm wavelength. In addition, the PC substrate can be designed and fabricated by adapting to any wavelengths from the visible to the near infrared. Thus, the SNSPDs fabricated on PC substrates can achieve high DE for vis–NIR wavelengths.

### Acknowledgments

This work was funded by the National Natural Science Foundation of China (Grant Nos. 61401441, and 61401443), Strategic Priority Research Program (B) of the Chinese Academy of Sciences (XDB04010200&XDB04020100), and National Basic Research Program of China (2011CBA00202).